\documentclass[10pt,onecolumn]{IEEEtran}
\usepackage[numbers,sort&compress]{natbib}
\usepackage{amsmath,amssymb,epsfig}
\usepackage{fancybox}
\usepackage{tikz}
\newtheorem{theorem}{\bf Theorem}
\newtheorem{corollary}{\bf Corollary}
\newtheorem{lemma}{\bf Lemma}

\begin{document}

\bibliographystyle{ieeetr}

\setlength{\parindent}{1pc}

\title{Loss Rate Estimators  and Properties for the Tree Topology}
\author{Weiping~Zhu  \thanks{Weiping Zhu is with University of New South Wales, Australia, email w.zhu@adfa.edu.au}}
\date{}
\maketitle

\begin{abstract} A large number of explicit estimators are proposed  in this paper for loss rate estimation in a network of the tree topology. All of the estimators are proved to be unbiased and consistent instead of asymptotic unbiased as that obtained in \cite{DHPT06} for a specific estimator. In addition,  a set of formulae are derived for the variances of various maximum likelihood estimators that unveil the connection between the path of interest and the subtrees connecting the path to observers. Using the formulae, we are able to not only rank the estimators proposed so far, including those proposed in this paper, but also identify the errors made in previous works. More importantly, using the formulae we can easily identify the most efficient explicit estimator from a pool that makes model selection feasible in loss tomography.
\end{abstract}


\section{Introduction}
\label{section1}

  Loss tomography has been studied for a number of years and a large number of estimators have been proposed for the networks of the tree topology \cite{CDHT99, CDMT99, CDMT99a, CN00, XGN06, BDPT02,ADV07, DHPT06, ZG05, GW03}. Among the proposed estimators, almost all of them rely on an iterative procedure, such as  the expectation and maximization (EM) or the Newton-Raphson algorithm, to approximate the solution of a likelihood equation that can be a high degree polynomial. Using approximation to solve a high degree polynomial has been widely criticised  for its computational complexity that increases with the number of descendants attached to the link or path to be estimated \cite{CN00}. Because of this, there has been a persistent interest in the research community to find an explicit estimator that performs as good as the iterative approach.  Apart from explicit estimators,  there are other issues in loss tomography that have not be solved. One of the issues is the theoretical variance of the estimates obtained by a maximum likelihood estimator (MLE). As far as we are aware of, there has been no a creditable result reported in a general form for the variances of MLEs although some expressions were presented and used in literature,  e.g.\cite{DHPT06}, that were obtained under specific conditions or assumptions. Because of this, the expressions cannot be used to evaluate the performance of an estimator. This paper is devoted to address these two issues and provide positive answers to them.

  There have been a number of attempts to propose explicit estimators and all of them aim at estimating the pass rate of a path, not a link,
  connecting the root of a network in the tree topology to an internal node of the network since there is a bijection between the pass rates of the paths and the pass rate of the links in a network of the tree topology.  However, due to the lack of a clear strategy in the search of explicit estimators,  all of the attempts are preliminary and produce little theoretical result. Apart from this, some of the results reported from the previous works, including those presented in \cite{DHPT06, ADV07, Zhu11a}, are incorrect or incomplete because of the lack of understanding the nature of the estimation or the use of unrealistical assumptions.

  To complete the analyses and correct those mistakes stated above,   we have undertaken a thorough and systematic investigation of the estimators proposed for loss tomography that aims at identifying the statistical principle and strategies that have been used or can be used in the tree topology. As a result, a number of findings are unveiled that show all of the estimators proposed previously  rely on observed correlations to infer the pass rates. The most popular strategy is to use all of the correlations available in estimation, such as the maximum likelihood estimator (MLE)  proposed in \cite{CDHT99}, that directly results in the use of high degree polynomials as the likelihood equations. Nevertheless, the qualities of the correlations, measured by the fitness between a correlation and the corresponding observation, are different, some are more fit than others.  Rather than using all of the correlations available but using a small portion of the high-quality correlations, we can have an explicit estimator that performs at least as good as the MLE proposed in \cite{CDHT99}.  This paper is devoted to those findings that contributes
to loss tomography in four-fold.
\begin{itemize}
\item A large number of explicit estimators are proposed on the basis of composite likelihood \cite{Lindsay88} that select some correlations from the available ones to estimate loss rates. The estimators are divided into three groups: the block wised estimators (BWE), the reduce scaled  estimators (RSE), and the individual based estimators (IBE).
\item The estimators in the three groups are proved to be unbiased rather than  asymptotic unbiased as that proved in   \cite{DHPT06}. A set of formulae are derived for the variances of the estimators in RSE and IBE, plus the MLE proposed in \cite{CDHT99}. The formulae show the variance of a loss rate estimator can be exactly expressed by the pass rate of the path of interest and the pass rate of the subtrees connecting the path of interest to the observers of interest.  The formulae also show the weakness of the result obtained in \cite{DHPT06} by the delta method.
\item  The efficiency of the estimators in IBE are compared with each other on the basis of the Fisher information that  shows an estimator considering  a correlation involving a few observers is more efficient than that considering a correlation involving many. Therefore, the estimator proposed in \cite{DHPT06} is the least efficient one. A similar conclusion is obtained for the estimators in BWE.
\item  Using the formulae, we able to identify an efficient estimator by examining the end-to-end pass rates that makes model selection not only possible but also feasible. A number of simulations are conducted to verify this finding.
    \end{itemize}

The rest of the paper is organised as follows. In Section \ref{related work},  we briefly introduce the previous works about explicit loss rate estimators and point out the weakness of them. In Section \ref{section2}, the loss model, the notations, and the statistics used in this paper are presented. In Section \ref{section3}, we  derive the MLE considering all available correlations for the networks of the tree topology. We then decompose the MLE into a number of components according to correlations and derive a number of likelihood equations for the components in Section \ref{section 4}. A statistic analysis of the proposed estimators is presented in Section \ref{section5} that details the statistical properties of the proposed estimators, one of them is the formulae to calculate the variances of various estimators. Given the large number of estimators, model selection is introduced in Section \ref{section6}. A strategy based on the formulae proposed is presented and a number of simulations are conducted that verify the feasibility of the proposed strategy.
Section \ref{section7} is devoted to concluding remark.
\section{Related Works}\label{related work}
Multicast Inference of Network Characters (MINC) is the pioneer of using the ideas proposed in \cite{YV96} into practice, where a
Bernoulli model is used to model  the loss behaviors of a link. Using
this model, the authors of \cite{CDHT99} derive an estimator to estimate the pass
rate of a path connecting the source to a node. The estimator is expressed
in a polynomial that is one degree less than the number of descendants
of the node \cite{CDHT99, CDMT99,CDMT99a}.  To ease the
concern of using numeric method to solve a higher degree polynomial $(
> 5 )$, the authors of \cite{DHPT06} propose an explicit estimator and claim the estimator has the same asymptotic variance as that obtained by the estimator proposed in \cite{CDHT99} to first
order. However, the claim is questionable because there has been no result about the variance of an estimator, including the MLE proposed in \cite{CDHT99}, and the result is obtained in \cite{DHPT06} is based on a unrealistical assumption, i.e. the loss rate of a link is very small. Under the assumption, almost all of the estimators proposed so far can achieve the same or better performance than that proposed in \cite{DHPT06}. In addition, the variance of the MLE used in the comparison in \cite{DHPT06} is also unrealistical because such a variance can only be obtained either by direct measurement or by letting the pass rate of the subtree rooted at the end of the path being estimated equal to or approach to 1.

In contrast to \cite{DHPT06}, \cite{ADV07, Zhu11a} propose an estimator that converts a general tree into a binary one and subsequently makes the likelihood equation into a quadratic equation of $A_k$  that is solvable analytically.   Experiments show the estimator preforms better than that in \cite{DHPT06} since the estimator uses more information in estimation. However, except experimental results, there is little theoretical analysis to demonstrate why it is better than that proposed in \cite{DHPT06}. In addition, although the author of \cite{Zhu11a} proves the estimator is a MLE, it is not clear that the MLE proposed in \cite{Zhu11a} is the same as that proposed in \cite{CDHT99}.

\section{Assumption, Notation and Sufficient Statistics} \label{section2}

 To make the following discussion simple and rigorous, we need to use  a large number of symbols that may overwhelm the readers who are not familiar with loss tomography. To assist them, the symbols  will be gradually introduced through the paper, where the frequently used symbols will be introduced in the following two sections and the others will be brought up later until needed. In addition, the most frequently used symbols and their meanings are presented in Table \ref{Frequently used symbols and description} for quick reference.

\subsection{Assumption}
To make loss tomography possible, probing packets, called probes, are multicasted from a source or a number of sources located on one side of a network to a number of receivers located on the other side of the network, where the paths connecting the sources to the receivers, via some routers, cover the links of interest. Statistical inference relies on the network topology and the correlation observed by the receivers to estimate the pass rate of the path shared by the paths from the source to the receivers. If a network that does not support multicast, unicast-based multicast can be used to achieve the same effect as multicast  \cite{HBB00},
\cite{CN00}. If the probes sent from sources to receivers are
far apart and network traffic remains statistically stable during the measurement, the observations are considered to be independent identical distributed
($i.i.d.$). In addition to probing, the losses occurred on a link or between links are assumed to be $i.i.d$  as well.

\subsection{Notation}\label{treenotation}

Let $T=(V, E)$
be the multicast tree used to dispatch probes from a source to a number of receivers, where  $V=\{v_0, v_1, ... v_m\}$ is a set of
nodes and $E=\{e_1,..., e_m\}$ is a set of directed links that connect the
nodes in $V$. By default $v_0$ is the root node of the multicast tree to which the source is attached.  The set of leaf nodes $R, R \subset V$ represents all receivers attached to $T$. If $f(i)$ is used to denote
the parent of node $i$, there is a correspondence between nodes and links, where $e_i$ is the link connecting $v_{f(i)}$ to
$v_i$. For instance, $e_1$ is the link connecting the parent of $v_1$, i.e. $v_0$, to $v_1$.

A multicast tree can be decomposed into a number of
 multicast subtrees at each of the internal nodes, where $T(i)$ denotes the subtree that has $e_i$ as its root link and $R(i)$ denotes the receivers attached to $T(i)$.  In addition, we use $d_i$  to denote the descendants attached to node $i$ that is a nonempty set if $i \notin R$. If $x$ is a set, $|x|$ denotes the number of elements in $x$ and $|d_i|$ denotes the number of descendants in  $d_i$. For example, Figure \ref{tree example} shows a complete binary multicast tree, where $R=\{v_8, v_9, \cdot\cdot, v_{15}\}$,  $R(2)=\{v_8, v_9, v_{10}, v_{11}\}$,  $d_{2}=\{4, 5\}$,  and $|d_{2}|=2$.

 If $n$ probes are sent from $v_0$ to $R$ in an experiment,
each of them gives rise of an independent realisation
of the passing (loss) process $X$. Let $X^{(i)}, i=1,...., n$ donate the $i-th$ process, where $x_k^i=1, k\in V$ if probe $i$
reaches $v_k$; otherwise $x_k^i=0$. The sample
$Y=(x_j^{(i)})^{i \in \{1,..,n\}}_{j \in R}$ comprises the observations of an experiment that can be divided into a number of sections according to $R(k)$, where $Y_k$ denotes the part of $Y$ obtained by $R(k)$.
If we use $y_j^i$ to denote the observation of receiver $j$ for probe $i$, we have $y_j^i=1$ if probe $i$ is observed by receiver $j$; otherwise, $y_j^i= 0$.

 Instead of using the loss rate of a link as the parameter to be estimated,  the pass rate of the path connecting $v_0$ to $v_k, k \in \{1,\cdot\cdot,m\}$ is used as the parameter and denoted by $A_k$. The empirical value of the parameter is equal to the number of probes arrived at node $k$ divided by the number of probes sent from the source, i.e. $n$. Given $A_k, k \in V\setminus v_0$, we are able to compute the pass rates of all links in $E$.  If $\alpha_k$ denotes the pass rate of link $k$ we have
 \begin{equation}
 \alpha_k=\dfrac{A_k}{A_{f(k)}}.
 \end{equation}
Given $\alpha_k$, $\bar{\alpha}_k=1-\alpha_k$ is the loss rate of link $k$.

\subsection{Statistics}

\label{mlesection}

To estimate  $A_k$ from end-to-end measurement, we need a likelihood function to connect the {\it i.i.d.}model defined previously to the observation obtained in an experiment. The MLE proposed previously considers  all of the probes observed by $R(k)$ that can be expressed as:
\begin{equation}
 n_k(d_k)=\sum_{i=1}^n \bigvee_{\substack{j \in R(k)}} y_j^i, \mbox{   }  k \in \{1,\cdot\cdot,m\}
 \label{nk1}
 \end{equation} \noindent that is the number of probes reaching node $k$ from the observation of $R(k)$, called the confirmed arrivals at node $k$.
To write a likelihood of $A_k$ for $n_k(d_k)$, $\beta_k$ and $\gamma_k$ are introduced to denote the pass rate of the subtrees rooted at node $k$ and the pass rate of the special multicast tree that connects $v_0$ to $R(k)$, via node $k$, respectively. Clearly $\gamma_k=A_k\cdot\beta_k,  k \in \{1,\cdot\cdot,m\}$ and $\hat\gamma_k=\dfrac{n_k(d_k)}{n}$ that is the empirical value of $\gamma_k$. Note that $\hat\gamma_j=\dfrac{n_j(j)}{n}, j \in R$ is the empirical pass rate of the path from the root to node $j$.

Given the assumptions made at the beginning of this section and the above definitions,  the likelihood function of $A_k$ for the observation obtained by $R(k)$, i.e. $n_k(d_k)$ can be created as follows:
\begin{equation}
P(A_k|Y)=(A_k\beta_k)^{n_k(d_k)}(1-A_k\beta_k)^{n-n_k(d_k)}.
\label{likelihood function}
\end{equation}
Given (\ref{likelihood function}), we can prove  $n_k(d_k)$ is a sufficient statistic with respect to ({\it wrt.}) the passing process of $A_k$ for the observation obtained by $R(k)$. Rather than using the well known factorisation theorem in the proof, we directly use the mathematic definition of a sufficient statistic (See definition 7.18 in \cite{RM96}) to achieve this. The definition {\it wrt.} the statistical model defined for the passing process is presented as a theorem here:

\begin{figure}
\begin{center}
\begin{tikzpicture}[scale=0.25,every path/.style={>=latex},every node/.style={draw,circle,scale=0.8}]
  \node            (b) at (25,30)  { $v_0$ };
  \node            (d) at (25,24) { $v_1$ };
  \node            (f) at (20,18)  { $v_2$ };
  \node            (g) at (30.5,18) { $v_3$ };
  \node            (j) at (16,12)  { $v_4$ };
  \node            (k) at (23,12) { $v_5$ };
  \node            (l) at (29,12) { $v_6$ };
  \node            (m) at (35,12)  { $v_7$ };
  \node            (r) at (12,6)  { $v_8$ };
  \node            (s) at (18,6) { $v_9$ };
  \node            (t) at (21,6) { $v_{10}$ };
   \node            (u) at (24.5,6)  { $v_{11}$ };
  \node            (v) at (27.5,6)  { $v_{12}$ };
  \node            (w) at (30.5,6) { $v_{13}$ };
  \node            (x) at (33.5,6) { $v_{14}$ };
  \node            (y) at (39,6) { $v_{15}$ };

  \draw[->] (b) edge (d);
  \draw[->] (d) edge (f);
  \draw[->] (d) edge (g);
  \draw[->] (f) edge (j);
  \draw[->] (f) edge (k);
  \draw[->] (g) edge (l);
  \draw[->] (g) edge (m);
  \draw[->] (j) edge (r);
  \draw[->] (j) edge (s);
  \draw[->] (k) edge (t);
  \draw[->] (k) edge (u);
  \draw[->] (l) edge (v);
  \draw[->] (l) edge (w);
  \draw[->] (m) edge (x);
  \draw[->] (m) edge (y);
\end{tikzpicture}
\caption{A Multicast Tree} \label{tree example}
\end{center}
\end{figure}
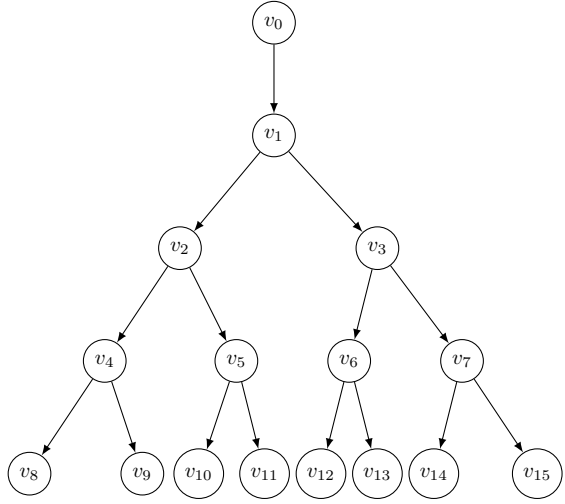

\begin{theorem}\label{complete minimal sufficient statistics}
Let $Y=\{X^{(1)},....,X^{(n)}\}$ be a random sample, governed by the
probability function $p_{A_k}(Y)$. The statistic $n_k(d_k)$ is minimal
sufficient for $A_k$ in respect of the observation of $R(k)$.
\end{theorem}

\begin{IEEEproof}
According to the definition of sufficiency, we need to prove
\begin{equation}
p_{A_k}(Y|n_k(d_k)=t)=\dfrac{p_{A_k}(Y)}{p_{A_k}(n_k(d_k)=t,Y)}
\label{suff-condition}
\end{equation} is independent of $A_k$.

Given (\ref{likelihood function}), the observation of $R(k)$ with $n_k(d_k)=t$ is a binomial distribution as follows
\[
p_{A_k}(n_k(d_k)=t)=\binom{n}{t}(A_k\beta_k)^{t}(1-A_k\beta_k)^{n-t}.
\]
Then, we have
\begin{eqnarray}
p_{A_k}(Y|n_k(d_k)=t)&=&\dfrac{(A_k\beta_k)^{t}(1-A_k\beta_k)^{n-t}}{\binom{n}{t}(A_k\beta_k)^{t}(1-A_k\beta_k)^{n-t}.
} \nonumber \\
=\dfrac{1}{\binom{n}{t}},
\end{eqnarray}
which is independent of $A_k$. Then, $n_k(d_k)$ is a sufficient statistic.

Apart from the sufficiency,
 $n_k(d_k)$,  as defined in (\ref{nk1}), is a count of the probes reaching $R(k)$ that counts each probe once and once only regardless of how many receivers observe the probe. Therefore,  $n_k(d_k)$ is a minimal sufficient statistic in regard to the observation of $R(k)$.
\end{IEEEproof}

\begin{table}[htdp]
\caption{Frequently used symbols and description}
\begin{center}
\begin{tabular}{|c|l|} \hline
Symbol & Desciption \\\hline
$T(k)$ & the subtree rooted at link $k$. \\ \hline
$d_k $& the descendants attached to node $k$. \\\hline
$R(k)$ & the receivers attached to $T(k)$. \\ \hline
$A_k$ & the pass rate of the path from $v_0$ to $v_k$. \\ \hline
$\beta_k$& the pass rate of the subtree rooted at node $k$. \\ \hline
$\gamma_k$& $A_k*\beta_k$, pass rate from $v_0$ to $R(k)$. \\ \hline
$x_k^i$ & the state of $v_k$ for probe $i$.  \\ \hline
$\sum_k$ & the $\sigma$-algebra created from $d_k$. \\ \hline
$n$ & the number of probes sent in an experiment, \\ \hline
$n_k(d_k)$ & the number of probes reaches $R(k)$. \\ \hline
$n_k(x)$ & the number of probes reaches the receivers attached to $T(j), j \in x$. \\ \hline
$I_k(x)$ & the number of probes observed by the members of $x$. \\ \hline
\end{tabular}
\end{center}
\label{Frequently used symbols and description}
\end{table}

\section{Problem Formulation and Analysis} \label{section3}

\subsection{Maximum Likelihood Estimator} \label{2.a}

Turning the likelihood function presented in (\ref{likelihood function}) into a log-likelihood function, we have
 \begin{equation}
 L(A_k|\Omega)=n_k(d_k)\log (A_k\beta_k)+(n-n_k(d_k))\log(1-A_k\beta_k).
 \label{likelihood}
 \end{equation}
 Differentiating (\ref{likelihood}) {\it wrt.} $A_k$
and letting the derivatives be 0, we have

\begin{eqnarray}
\dfrac{n_k(d_k)}{A_k}-\dfrac{(n-n_k(d_k))\beta_k}{1-A_k\beta_k}=0.
\label{likelihood equation}
\end{eqnarray}

\noindent  Given the {\it i.i.d.} model assumed previously and the multicast used in probing, we have the following equation to link the observation of $R(k)$ to $\beta_k$ that is defined as the pass rate of the subtree rooted at node $k$
\begin{equation}
1-\beta_k=\prod_{j \in d_k} (1-\dfrac{\gamma_j}{A_k}).
\label{beta-k}
\end{equation}
Solving $\beta_k$ from (\ref{beta-k}) and using it in (\ref{likelihood equation}), we have a likelihood equation as
\begin{equation}
1-\dfrac{n_k(d_k)}{n \cdot A_k}=\prod_{j \in d_k} (1-\dfrac{\gamma_j}{A_k}).
\label{realmle1}
\end{equation}
Using $\gamma_k$ to replace $\dfrac{n_k(d_k)}{n}$ since the latter is the empirical value of the former, we have a likelihood as follows:
\begin{equation}
1-\dfrac{\gamma_k}{A_k}=\prod_{j \in d_k} (1-\dfrac{\gamma_j}{A_k})
\label{minc}
\end{equation}
(\ref{minc}) is identical to the estimator proposed in \cite{CDHT99}.  In order to find the correlations considered in the MLE, we use (\ref{realmle1}) rather than (\ref{minc}) in the rest of this section because it explicitly connects observations to correlations.

\subsection{Correlations and  Observation}
%
%

 To find the number of correlations considered by the MLE, both sides, the right hand side (RHS) and the left hand side (LHS), of  (\ref{realmle1}) are expanded to show the correspondences between observations and correlations. The correlations are called  the {\it predictors} of the observations since the former predicates the latter.  For instance, $\gamma_i\cdot \gamma_j/A_k, i, j \in d_k \land i \neq j$ is the predictor of the portion of probes that are simultaneously observed by at least two receivers attached to subtree $i$ and subtree $j$, respectively.

To find the correlations involved in (\ref{realmle1}),
 a $\sigma$-algebra, $S_k$, is created over $d_k$. Let $\Sigma_k=S_k \setminus \emptyset$ be the non-empty sets in $S_k$, each of $\Sigma_k$ corresponds to a pair of predictor and observation.  If  the number of elements in a member of $\Sigma_k$ is defined as the degree of the correlation, $\Sigma_k$ can be divided into $|d_k|$ exclusive groups for
 correlations varying from 1 to $|d_k|$. Let $S_k(i), i \in \{1,\cdot\cdot,|d_k|\}$ denote the group that considers the correlation involving $i$ members of $d_k$,  we call it $i-$wise correlation. For example, if $d_k=\{i,j,k,l\}$, $S_k(2)=\{(i,j),(i,k),(i,l),(j,k),(j,l),(k,l)\}$ that consists of all of the pairwise correlations in $d_k$ and $S_k(3)=\{(i,j,k),(i,j,l),(i,k,l),(j,k,l)\}$ consists of the tripletwise correlations.

Given $\Sigma_k$,  $n_k(d_k)$ can be decomposed into the probes that are observed by the members of  $\Sigma_k$. If $x$ is a member of $\Sigma_k$ and $|x|>1$, a probe that is observed by $x$ is defined as if and only if at least a receiver attached to subtree $j, j \in x$ observes the probe, called simultaneous observation.
 To explicitly express $n_k(d_k)$ by $n_j(d_j), j \in d_k$,  $I_k(x), x \in \Sigma_k$ is introduced to return  the number of probes observed by  $x$ in an experiment.
If $u_j^i$ is the observation of $R(j)$ for probe $i$, which is equal to
\[
u_j^i=\bigvee_{k \in R(j)} y_k^i,
\]
 we have
\begin{equation}
I_k(x)=\sum_{i=1}^n \bigwedge_{j \in x} u_j^i, \mbox{\vspace{1cm} } x \in \Sigma_k.
\label{I-k x}
\end{equation}
If $x=(j)$,
\[
I_k(x)=n_j(d_j), j \in d_k,
\]
 Then, we have

\begin{equation}
n_k(d_k)=\sum_{i=1}^{|d_k|}(-1)^{i-1}\sum_{x \in S_k(i)} I_k(x). \label{n_k value}
\end{equation}
(\ref{n_k value}) states that $n_k(d_k)$ is equal to a series of alternating adding and subtracting operations that ensure each probe observed by $R(k)$ is counted once and once only in $n_k(d_k)$.

\subsection{Correlations in  MLE}

Given  (\ref{n_k value}), we are able to prove the MLE proposed in \cite{CDHT99} considers all of the correlations in $\Sigma_k$.

\begin{theorem} \label{minctheorem}

\begin{enumerate}
\item (\ref{realmle1}) is a full likelihood estimator that considers all of the correlations in $\Sigma_k$;
\item (\ref{realmle1}) consists of  observed values and their predictors, one for a member of $\Sigma_k$; and
\item the estimate obtained from (\ref{realmle1})  is a fit that minimises an alternating differences between observed values and corresponding predictors.
    \end{enumerate}
\end{theorem}
\begin{IEEEproof}
(\ref{realmle1}) is a full likelihood estimator that considers all of the correlations in $d_k$. To prove 2) and 3),  we expand the both sides of (\ref{realmle1}) to pair the observed values with the  predictors of them according to $S_k$. We take three steps to achieve the goal.
\begin{enumerate}
\item  If  we use (\ref{n_k value}) to replace $n_k(d_k)$ from LHS of (\ref{realmle1}),  the LHS becomes:
\begin{equation}
1-\dfrac{n_k(d_k)}{n\cdot A_k}= 1-\dfrac{1}{n\cdot A_k}\big
[\sum_{i=1}^{|d_k|}(-1)^{i-1}\sum_{x \in S_k(i)}I_k(x)].
\label{nkexpansion}
\end{equation}
\item If we expand the product term located on  the RHS of (\ref{realmle1}), we have:
\begin{equation}
\prod_{j \in d_k}(1-\dfrac{\gamma_j}{A_k})=1-\sum_{i=1}^{|d_k|}(-1)^{i-1}\sum_{x \in S_k(i)}\dfrac{\prod_{j \in x} \gamma_j}{A_k^i}. \label{prodexpansion}
\end{equation} where the alternative adding and subtracting operations intend to remove the impact of redundant observation in  $n_k(d_k)$.
\item Deducting 1 from both (\ref{nkexpansion}) and (\ref{prodexpansion})
and then multiplying the results by $A_k$, (\ref{realmle1}) turns to
\begin{eqnarray}
\sum_{i=1}^{|d_k|}(-1)^{i}\sum_{x \in S_k(i)}\dfrac{I_k(x)}{n}
=\sum_{i=1}^{|d_k|}(-1)^{i}\sum_{x \in S_k(i)}\dfrac{\prod_{j \in x} \gamma_j}{A_k^{i-1}}. \label{statequal}
\end{eqnarray}
It is clear there is a correspondence between the terms across
the equal sign, where the terms on the LHS are the observed values and the terms on the RHS are the predictors. If we rewrite
(\ref{statequal}) as
\begin{equation}
\sum_{i=1}^{|d_k|}(-1)^{i}\sum_{x \in S_k(i)}\Big(\dfrac{I_k(x)}{n} -\dfrac{\prod_{j \in x}
\gamma_j}{A_k^{i-1}}\Big)=0, \label{correspondence}
\end{equation}
\end{enumerate}
the correspondence between correlations and observed values becomes obvious in  (\ref{correspondence}).
\end{IEEEproof}

To distinguish the MLE from the others proposed in this paper, we call it original MLE in the rest of the paper.

\section{Explicit Estimators based on Composite Likelihood} \label{section 4}
(\ref{correspondence}) shows that the original MLE considers all of the correlations available in $\Sigma_k$ that makes the estimator a high degree polynomial if the number of subtrees rooted at node $k$ is larger than 6. To find explicit estimators in this circumstance, we must reduce the number of correlations considered by an estimator and use composite likelihood to create likelihood functions, composite likelihood is also called
pseudo-likelihood by Besag \cite{Besay74}. Three strategies are used here to reduce the number of correlations used in estimation: block-wised, reduce scaled, and individual based. The block-wised strategy divides all correlations into blocks, each consists of the correlations of the same degree, from pairwise to $d_k$-wise. The reduce scaled strategy, as named, reduces the number of subtrees considered in estimation. The individual based one considers each correlation separately.

\subsection{Block-wised Estimator (BWE)}

Let $\psi_k(x)=\prod_{j \in x} \beta_j$, where $x \subset d_k$, be the pass rate of the subtrees in $x$. If the number of probes reaching node $k$ is denoted by $\hat n_k(d_k)$, the empirical value of $\psi_k(x)$ is equal to
\[
\dfrac{I_k(x)}{\hat n_k(d_k)}.
\]
From (\ref{correspondence}),  $|d_k|-1$ block-wised likelihood functions can be identified, from pairwise likelihood to $|d_k|$-wise likelihood.  Each of them corresponds to an item in the first summation  of (\ref{correspondence}). In order to have a unique likelihood function for all of them, let the single-wise likelihood function be 1 and let the $i$-wise likelihood function be $L_c(i; A_k; y)$. Then, the block-wised likelihood functions can be expressed uniformly.

\begin{lemma} \label{recursive corollary}
There are a number of composite likelihood functions, one for a type of correlations, varying from pairwise to $|d_k|$-wise. The composite likelihood function  $L_c(i; A_k; y), i \in \{2,\cdot\cdot, |d_k|\}$ has a  form as follows:
\begin{eqnarray}
L_c(i; A_k; y) &=&\dfrac{\prod_{x \in S_k(i)}(A_k\psi_k(x))^{n_k(x)}(1-A_k\psi_k(x))^{n-n_k(x)}}{\prod_{y \in S_k(i-1)}(A_k\psi_k(y))^{n_k(y)}(1-A_k\psi_k(y))^{n-n_k(y)}}. \nonumber \\
&& i \in \{2,\cdot\cdot,|d_k|\}
\label{recursive form}
\end{eqnarray}
 \end{lemma}
\begin{IEEEproof}
The nominator on the RHS of (\ref{recursive form}) is the likelihood function considering the correlations from pairwise to $i$-wise inclusively and the denominator is the likelihood functions from single-wise to $(i-1)$-wise. The quotient of them is the likelihood dedicated to the $i$-wise correlation.
\end{IEEEproof}
 Let $A_k(i)$ be the estimator derived from $L_c(i; A_k; y)$. Then,  we have the following theorem.
\begin{theorem} \label{all explicit}
Each of the composite likelihood
equations obtained from (\ref{recursive form}) is an explicit estimator of $A_k$ that is as follows:
\begin{equation}
A_k(i)=\Big (\dfrac{\sum_{\substack{ x \in S_k(i)}}
\prod_{j \in x} \gamma_j}{\sum_{x \in S_k(i)}\dfrac{I_k(x)}{n}}{\Big )} ^{\frac{1}{i-1}},  i \in
\{2,.., |d_k|\}. \label{approximateestimator}
\end{equation}
\end{theorem}
\begin{IEEEproof}
Firstly, we rewrite (\ref{recursive form}) into a log-likelihood function. We then differentiate the log-likelihood function {\it wrt} $A_k$ and let the derivative  be 0. The likelihood equation as (\ref{approximateestimator}) follows.
\end{IEEEproof}
In the rest of the paper, $A_k(i)$ is used for the $i-wise$ estimator and $\widehat A_k(i)$ for the estimate obtained by $A_k(i)$.

\subsection{Reduce Scaled Estimator (RSE)}

Instead of grouping the correlations of the same degree into a likelihood equation, the correlations can be grouped according to the subtrees rooted at node $k$. Since $x, x \subset d_k$ are selected, the estimators are called RSE. The log-likelihood function of the correlations within $x$ is as follows:
 \begin{equation}
 L(A_k|\Omega_x)=n_k(x)\log (A_k\beta_k(x))+(n-n_k(x))\log(1-A_k\beta_k(x))
 \label{likelihood RSE}
 \end{equation}
 where $n_k(x)$ is the number of probes reaching node $k$ from the observations of the receivers attached to $T(j), j \in x$ that equals to
\begin{equation}
n_k(x)=\sum_{i=1}^{|x|}(-1)^{i-1}\sum_{y \in S_k(x)} I_k(y) \nonumber
\end{equation}
where $S_k(x)$ is the $\sigma$-algebra created over $x$.
 $\beta_k(x)$ is the pass rate of  $T(j), j \in x$, and defined as
\begin{equation}
1-\beta_k(x)=\prod_{j \in x} (1-\dfrac{\gamma_j}{A_k}).
\label{beta-k1}
\end{equation}
Then, a similar likelihood equation as (\ref{realmle1}) is obtained and presented as follows:
\begin{equation}
1-\dfrac{n_k(x)}{n \cdot A_k}=\prod_{j \in x} (1-\dfrac{\gamma_j}{A_k}).
\label{estimator MLEPC1}
\end{equation}
Clearly, (\ref{estimator MLEPC1}) is a reduce scaled of (\ref{realmle1}). If $|x|<5$, the equation is solvable. The estimator is denoted by $Am_k(x)$.

\subsection{Individual based Estimator (IBE)}

 The likelihood functions of the estimators in IBE have a similar structure as (\ref{likelihood RSE}), where $\beta_k(x)$ and $n_k(x)$ are replaced by $\psi_k(x)$ and $I_k(x)$, respectively.  Let $\Sigma_k'=\Sigma_k \setminus S_k(1)$ be the correlations considered by IBE. Then, the log-likelihood function for $A_k$ given observation $I_k(x)$ is equal to
 \begin{eqnarray}
 L(A_k|I_k(x))=I_k(x)\log (A_k\psi_k(x))+(n-I_k(x))\log(1-A_k\psi_k(x)),  \mbox{   } x \in \Sigma_k'.
 \label{Al likelihood1}
 \end{eqnarray}
 We then have the following theorem.
\begin{theorem}  \label{local estimator}
Given (\ref{Al likelihood1}), $A_k\psi_k(x)$ is a Bernoulli process. The MLE for $A_k$ given $I_k(x)$ equals to
\begin{equation}
Al_k(x)=\Big(\dfrac{\prod_{j\in x} \gamma_j}{\dfrac{I_k(x)}{n}}\Big)^{\frac{1}{\|x|-1}}.  \mbox{   }    x \in \Sigma_k'
\label{local estimator1}
\end{equation}
\end{theorem}
	\begin{IEEEproof} Using the same procedure as that used in \ref{2.a}, we have the theorem.
\end{IEEEproof}

Comparing (\ref{approximateestimator}) with (\ref{local estimator1}), we can find that $\widehat Al_k(x)$, where $\|x|=i$, is a type of geometric mean and $\widehat A_k(i)$ is the arithematic mean of $\widehat Al_k(x), x \in S_k(i)$. Therefore, $A_k(i)$ is more robust than $Al_k(x)$.

Using and combining the strategies presented above, we can have various explicit estimators. In fact, the estimator proposed in \cite{ADV07, Zhu11a} is one of them that divides $d_k$ into two groups and only considers the pairwise correlations between the members of the two groups. Therefore, although the estimator proposed in \cite{ADV07, Zhu11a} is a MLE in terms of the observation used in estimation, it is not the same as (\ref{minc}).

\section{Properties of the Estimators} \label{section5}

To evaluate the performance of the estimators proposed in this paper, we need to study the statistical properties of them, i.e., unbiasedness, consistency,  uniqueness, variance, and efficiency. This section is devoted to the properties that consist of a number of  lemmas, theorems and corollaries.

\subsection{Unbiasedness and Consistency}

The original MLE has been proved to be unbiased and consistent in \cite{CDHT99}. Using the same methodology, we are able to prove the unbiasedness and consistency of
the estimators in RSE. Thus, our attention here is focused on the properties of the estimators in IBE and BWE.

For the unbiasedness of $Al_k(x), x \in \Sigma_k'$, we have the following theorem.

 \begin{theorem} \label{local maximum}
$Al_k(x)$ is a unbiased estimator.
\end{theorem}

\begin{IEEEproof}
Let $\hat n_k(d_k)$ be the number of probes reaching $v_k$ and let $z_j, j \in d_k$ be the pass rate of $T(j)$. In addition, let  $\overline{z_j}=\frac{n_j(d_j)}{\hat n_k(d_k)}$ be the sample mean of $z_j$ and $\overline{A_k}=\frac{\hat n_k(d_k)}{n}$ be the sample mean of $A_k$. Note that $z_j$ and $z_l, j, l \in d_k$ are independent from each other if $ j \neq l$. Apart from those, we use $x_k^i\prod_{j \in x} z_j$ to replace $\bigwedge_{j \in x} y_j^i$ in the following derivation since the latter is equal to $\prod_{j\in x} y_j^i$ which is equal to $x_k^i\prod_{j \in x} z_j$. We then have
\begin{eqnarray}
E(\widehat Al_k(x)^{\|x|-1})&=&E\Big(\dfrac{\prod_{j\in x} \hat\gamma_j}{\dfrac{I_k(x)}{n}}\Big) \nonumber \\
&=& E\Big(\dfrac{\prod_{j\in x} \dfrac{n_j(d_j)}{n}}{\dfrac{\sum_{i=1}^n \bigwedge_{j \in x} y^i_j}{n}} \Big )\nonumber \\
&=& E\Big(\dfrac{ (\dfrac{\hat n_k(d_k)}{n})^{\|x|}\prod_{j\in x} \dfrac{n_j(d_j)}{\hat n_k(d_k)}}{\dfrac{\hat n_k(d_k)}{n} \dfrac{\sum_{i=1}^{\hat n_k(d_k)}\prod_{j \in x} z_j}{\hat n_k(d_k)}} \Big ) \nonumber \\
&=& E\Big((\dfrac{\hat n_k(d_k)}{n})^{\|x|-1}\Big)E\Big(\dfrac{\prod_{j\in x} \dfrac{1}{\hat n_k(d_k)}\sum_{i=1}^{\hat n_k(d_k)} {z_j}}{\sum_{i=1}^{\hat n_k(d_k)} \dfrac{1}{\hat n_k(d_k)}\prod_{j \in x} {z_j}}\Big ) \label{second element}\\
&=&E\Big (\overline{A_k}^{\|x|-1}\Big)
\label{weighted mean}
\end{eqnarray}
The theorem follows.
\end{IEEEproof}
Given theorem \ref{local maximum}, we have the follow corollary.
 \begin{corollary} \label{global expect}
 $A_k(i)$ is a unbiased estimator.
 \end{corollary}
    \begin{IEEEproof}
According to theorem \ref{local maximum}, we have
\begin{eqnarray}
E(\widehat A_k(i))&=&E\Big(\overline{A_k}\Big)E\Big(\big(\dfrac{\sum_{x \in S(i)}\prod_{j\in x} \overline{z_j}}{\sum_{x \in S(i)}\prod_{j\in x}  z_j}\big)^{\frac{1}{i-1}} \Big)\nonumber \\
&=&E\Big (\overline{A_k}\Big)
\label{global estimate}
\end{eqnarray}
\end{IEEEproof}

%
%
%
 Note that we here prove that  $Al_k(x), x \in \Sigma_k'$ are unbiased estimators, rather than asymptotic unbiased ones as that obtained in \cite{DHPT06} for $Al_k(d_k)$.

Further, we can
prove $\widehat Al_k(x), \|x|=i$ and $\widehat A_k(i)$ are consistent estimates in the following lemma and theorem.

\begin{lemma} \label{two proofs}
$\widehat Al_k(x)$ is a consistent estimate of $A_k$.
\end{lemma}
\begin{IEEEproof}
We have the following two points to prove the lemma.
\begin{enumerate}
\item Theorem \ref{local maximum} shows that $\widehat Al_k(x)$ is equivalent to the mean of $A_k$. Then, according to the law of large number, $\widehat Al_k(x)\rightarrow A_k$. \label{first proof}
    \item From the above and the continuity of $Al_k(x)$ on the values of $\gamma_j, j \in x$ and $I_k(x)/n$ generated as $A_k$ ranges over its support set, the result follows.
        \end{enumerate}
\end{IEEEproof}
Then, we have
\begin{theorem} \label{approximate consistent}
   $\widehat A_k(i)$ is a consistent estimate of $A_k$.
\end{theorem}
\begin{IEEEproof}
  As stated,  $\widehat A_k(i)$ is a mean of $\widehat Al_k(x), x \in S_k(i)$ that satisfies the followings inequality
\begin{equation}
\min_{x \in S_k(i)}  \widehat Al_k(x)^{\frac{1}{i-1}} \leq \widehat A_k(i) \leq  \max_{x \in S_k(i)} \widehat Al_k(x)^{\frac{1}{i-1}}.
\label{average mean1}
\end{equation}
Since all of $Al_k(x), x \in S_k(i)$ are consistent estimators, $A_k(i)$ is a consistent estimator.
\end{IEEEproof}

For the uniqueness of $A_k(i)$, we have.
\begin{theorem}
If
\[
\sum_{\substack{ x \in S_k(i)}} \prod_{j \in x} \hat\gamma_j < \sum_{x \in S_k(i)}\dfrac{I_k(x)}{n},
\]
there is only one solution in $(0,1)$ for $\widehat A_k(i), 2 \leq i \leq |d_k|$.
\end{theorem}
\begin{IEEEproof}
Since the support of $ A_k$ is in (0,1), we can reach this conclusion from (\ref{approximateestimator}).
\end{IEEEproof}

\subsection{Efficiency and Variance of  $Al_k(x)$, $Am_k(x)$, and the original MLE}
Given (\ref{Al likelihood1}), we have the following theorem for the Fisher information of an observation, $y$, in IBE that can be used to determine the efficiency of $Al_k(x), x \in \Sigma_k'$.
\begin{theorem} \label{Al fisher}
The Fisher information of $y$ on $Al_k(x), x \subset d_k$ is equal to $ \dfrac{\psi_k(x)}{A_k (1-A_k \psi_k(x))}$.
\end{theorem}
\begin{IEEEproof}
Considering $I_k(x)=y$ is the observation of the receivers attached to $x$, we have the following as the likelihood function of the observation:
\begin{equation}
L(A_k|y)=y\log (A_k\psi_k(x))+(1-y)\log(1-A_k\psi_k(x)).
\label{Al likelihood for single}
\end{equation}
Differentiating (\ref{Al likelihood for single}) {\it wrt} $A_k$, we have
\begin{eqnarray}
\dfrac{\partial L(A_k|y)}{\partial A_k}=\dfrac{y}{A_k}-\dfrac{(1-y)\psi_k(x)}{1-A_k\psi_k(x)}
\end{eqnarray}
We then have
\begin{eqnarray}
\dfrac{\partial^2 L(A_k|y)}{\partial A_k^2}&=& -\dfrac{y}{A_k^2}-\dfrac{(1-y)\psi_k(x)^2}{(1-A_k\psi_k(x))^2}
\end{eqnarray}
If ${\cal I}(Al_k(x)|y)$ is used to denote the Fisher information of observation $y$ for $A_k$ in $Al_k(x)$, we  have
\begin{eqnarray}\label{fisher}
{\cal I}(Al_k(x)|y)&=&-E(\dfrac{\partial^2 L(A_k|y)}{\partial A_k^2}) \nonumber \\
&=&\dfrac{E(y)}{A_k^2}+\dfrac{E(1-y)\psi_k(x)^2}{(1-A_k\psi_k(x))^2} \nonumber \\
&=&\dfrac{\psi_k(x)}{A_k (1-A_k \psi_k(x))}
\end{eqnarray}
that is the information provided by $y$ for $A_k$.
\end{IEEEproof}
Given (\ref{fisher}), we are able to have a formula for the Fisher information of the original MLE and the estimators in RSE. In order to achieve this, let $\beta_k(d_k)=\beta_k$. Then, we have the following corollary.
\begin{corollary} \label{MLE fisher}
The Fisher information of observation $y$ for $A_k$ in the original MLE and $Am_k(x), x \subset d_k$ is equal to
\begin{equation}
 \dfrac{\beta_k(x)}{A_k (1-A_k \beta_k(x))}, \mbox{     } x \subseteq d_k.
 \label{MLE fisher equ}
 \end{equation}
\end{corollary}
\begin{IEEEproof}
Replacing $n_k(d_k)$ and $n_k(x)$ by $y$ and replacing $n-n_k(d_k)$ and $n-n_k(x)$ by $1-y$ from (\ref{likelihood}) and (\ref{likelihood RSE}), respectively, and then using the same procedure as that used in theorem \ref{Al fisher} on the log-likelihood functions, the corollary follows.
\end{IEEEproof}
Because of the similarity between (\ref{fisher}) and (\ref{MLE fisher equ}), the two equations have the same features in terms of support, singularity, and maximums. After
eliminating the singular points of them, the support of $A_k$ is in $(0,1)$ and the support of $\beta_k(x)$ (or $\psi_k(x)$) is in $[0, 1]$.
Both (\ref{fisher}) and (\ref{MLE fisher equ}) are convex functions in the support and reaches the maximum at the points of  $A_k \rightarrow 1, \beta_k(x) =1$ (or ($\psi_k(x)=1$) and $A_k\rightarrow 0, \beta_k(x)=1$ (or ($\psi_k(x)=1$).
Given $A_k$, (\ref{MLE fisher equ}) is  a monotonic increase function of $\beta_k(x)$ whereas (\ref{fisher}) is a a monotonic increase function of $\psi_k(x)$. Despite the similarity, the efficiency of $Al_k(x)$ and $Am_k(x)$ go to opposite direction if $x$ is replaced by $y$,  $x \subset y $. $Al_k(y)$ is less efficient than $Al_k(x)$ since $\psi_k(x) > \psi_k(y)$, but $Am_k(y)$ is more efficient than $Am_k(x)$ since $\beta_k(x) < \beta_k(y)$.
Given theorem \ref{Al fisher}, we are able to compare the efficiency of $Al_k(x), x \in \Sigma_k'$ and have the following corollary.
\begin{corollary}
The efficiency of $Al_k(x), x \in \Sigma_k'$ forms a partial order that is the same as that formed on the inclusion of the members in $\Sigma_k'$, where  the most efficient estimator  must be one of the $Al_k(x), x \in S_k(2)$ and  the least efficient  one must be $Al_k(d_k)$.
\end{corollary}
\begin{IEEEproof}
The inclusion in $\Sigma_k'$ forms a partial order, where a member of $S_k(i)$ is included by
at least one in $S_k(i+1), i+1 \leq |d_k|$. Because all of  the members in $\Sigma_k'$ except those in $S_k(2)$ include  at least one of $S_k(2)$,  the most efficient estimator must be one of $Al_k(x), x \in S_k(2)$. On the other hand, $\forall x  \{x \in \Sigma_k' \rightarrow x \subseteq d_k\}$, $Al_k(d_k)$ is the least efficient estimator in $Al_k(x), x \in \Sigma_k'$.
\end{IEEEproof}

Equation (\ref{minc}),  $Am_k(x)$, and $Al_k(x)$ are of MLEs that have different focuses on the observations obtained. Because of this, they share a number of features, including likelihood functions and efficient equations. In addition, the variances of them can be expressed by a general function. Let $mle$ denote all of them. Then, we have a theorem for the variances of the estimators in $mle$.

\begin{theorem} \label{Al variance}
The variance of the estimators in $mle$ equal to
\begin{equation}
var(mle)=\dfrac{A_k (1-A_k\delta_k(x) )}{\delta_k(x)}, \mbox{  } x \subseteq d_k
\label{Al variance1}
\end{equation}
where $\delta_k(x)$ is the pass rate of the subtrees in $x$ that is calculated on the basis of the definition of  individual estimators.
\end{theorem}
\begin{IEEEproof}
The passing process described by (\ref{Al likelihood1}) is a Bernoulli process that falls into the exponential family and satisfies the regularity conditions presented in \cite{Joshi76}. Thus, the variance of  an estimator in $mle$ reaches the Cram\'{e}r-Rao bound that is the reciprocal of the Fisher information.
\end{IEEEproof}
(\ref{Al variance1}) unveils such a fact that the estimates obtained by an estimator spread out more widely than that obtained by direct measurement. The wideness is determined by $\delta_k(x)$, the pass rate of the subtrees connecting node $k$ to observers. If $\delta_k(x)=1$, there is no further spread-out than that obtained by direct measurement. Otherwise, the variance estimated increases as the decreases of $\delta_k$ and in a super linear fashion.

\subsection{Efficiency and Variance of BWE}

 The estimate obtained by $A_k(i)$ is a type of the arithmetic mean of $\widehat Al_k(x), x \in S_k(i)$ that has the same advantages and disadvantages as the arithmetic mean. $A_k(i)$ differs to $\widehat Al_k(x)$ by using a statistic that is not sufficient since some probes are considered more than once. Because of this, the Fisher information cannot be used to evaluate the efficiency of an estimator in BWE.
 Nevertheless, as a special arithmetic mean of $Al_k(x), |x|=i$, $A_k(i)$ shares many features as $Al_k(x)$.
Thus, $A(i)$ is more efficient than $A(i+1)$ and the variance of $A(i)$ is smaller than that of $A(i+1)$.


\section{Model Selection and Simulation} \label{section6}
The large number of estimators in IBE, RSE and BWE, plus the original MLE, make model selection possible. However, to find the most suitable one in terms of efficiency and computational complexity is a hard task since the two goals conflict each other. Although one is able to identify the the most suitable estimator by computing the Kullback-Leigh divergence or the composite Kullback-Leigh divergence of the estimators, the cost of computing the Akaike information criterion (AIC) for each of the estimators makes this approach prohibitive.  Nevertheless, the derivation of (\ref{MLE fisher equ}) successfully solves the problem since (\ref{MLE fisher equ}) shows the most suitable estimator must have the subtrees that have the highest end-to-end pass rates.

\begin{table*}[th]
  \centering
  \scriptsize
  \begin{tabular}{|l|r|r|r|r|r|r|r|r|r|r|}  \hline
 Estimators &\multicolumn{2}{|c|}{Full Likelihood} & \multicolumn{2}{|c|}{$A_k(2)$} & \multicolumn{2}{|c|}{$A_k(3)$} &\multicolumn{2}{|c|}{$Al_k(x), |x|=2$} & \multicolumn{2}{|c|}{$Al_k(x), |x|=3$}	\\ \hline
samples & Mean & Var &	Mean & Var &	Mean & Var	& Mean & Var &  Mean &	Var	\\ \hline
300&	0.0088&	1.59E-05&	0.0088&	1.59E-05&	0.0088&	1.64E-05&	0.0087&	 1.59E-05&	0.0087&	1.61E-05 \\ \hline
600&	0.0089&	1.12E-05&	0.0089&	1.12E-05&	0.0089&	1.13E-05&	0.0089&	 1.10E-05&	0.0088&	1.12E-05 \\ \hline
900&	0.0092&	7.76E-06&	0.0092&	7.82E-06&	0.0091&	7.84E-06&	0.0092&	 7.90E-06&	0.0092&	8.15E-06 \\ \hline
1200&	0.0095&	6.13E-06&	0.0095&	6.13E-06&	0.0094&	6.17E-06&	0.0095&	 6.16E-06&	0.0095&	5.97E-06 \\ \hline
1500&	0.0096&	4.55E-06&	0.0096&	4.55E-06&	0.0096&	4.80E-06&	0.0096&	 4.78E-06&	0.0096&	4.33E-06 \\ \hline
1800&	0.0096&	1.82E-06&	0.0096&	1.81E-06&	0.0096&	1.92E-06&	0.0097&	 1.92E-06&	0.0096&	1.90E-06 \\ \hline
2100&	0.0097&	3.14E-06&	0.0097&	3.11E-06&	0.0097&	3.14E-06&	0.0097&	 3.02E-06&	0.0097&	3.08E-06 \\ \hline
2400&	0.0100&	1.32E-06&	0.0100&	1.32E-06&	0.0100&	1.36E-06&	0.0100&	 1.29E-06&	0.0099&	1.28E-06 \\ \hline
2700&	0.0100&	1.72E-06&	0.0100&	1.72E-06&	0.0100&	1.74E-06&	0.0100&	 1.81E-06&	0.0100&	1.83E-06 \\ \hline
3000&	0.0102&	2.96E-06&	0.0102&	2.97E-06&	0.0102&	3.01E-06&	0.0102&	 3.04E-06&	0.0102&	2.95E-06 \\ \hline
4800&	0.0103&	1.74E-06&	0.0103&	1.74E-06&	0.0103&	1.74E-06&	0.0103&	 1.75E-06&	0.0103&	1.81E-06 \\ \hline
9900&	0.0099&	8.18E-07&	0.0099&	8.23E-07&	0.0099&	8.20E-07&	0.0099&	 8.05E-07&	0.0099&	8.60E-07 \\ \hline
\end{tabular}
  \caption{Simulation Result of a 8-Descendant Tree with Loss Rate=$1\%$}
  \label{Tab2}
\end{table*}

\begin{table*}
\centering
\scriptsize
\begin{tabular}{|l|r|r|r|r|r|r|r|r|r|r|}  \hline
 Estimators &\multicolumn{2}{|c|}{Full Likelihood} & \multicolumn{2}{|c|}{$A_k(2)$} & \multicolumn{2}{|c|}{$A_k(3)$} &\multicolumn{2}{|c|}{$Al_k(x), |x|=2$} & \multicolumn{2}{|c|}{$Al_k(x), |x|=3$}	\\ \hline
samples & Mean & Var &	Mean & Var &	Mean & Var	& Mean & Var &  Mean &	Var	\\ \hline
300&	0.0088&	1.59E-05&   0.0089&	1.64E-05&	0.0089&	1.68E-05&	0.0091&	 2.36E-05&	0.0088&	1.95E-05 \\ \hline
600&	0.0089&	1.12E-05&	0.0089&	1.14E-05&	0.0089&	1.16E-05&	0.0088&	 1.46E-05&	0.0089&	1.26E-05 \\ \hline
900&	0.0091&	7.76E-06&	0.0091&	7.80E-06&	0.0091&	7.83E-06&	0.0092&	 9.74E-06&	0.0091&	8.67E-06 \\ \hline
1200&	0.0094&	6.13E-06&	0.0094&	6.16E-06&	0.0094&	6.18E-06&	0.0096&	 7.09E-06&	0.0095&	6.16E-06 \\ \hline
1500&	0.0096&	4.55E-06&	0.0096&	4.72E-06&	0.0096&	4.81E-06&	0.0097&	 4.36E-06&	0.0096&	4.45E-06 \\ \hline
1800&	0.0096&	1.82E-06&	0.0096&	1.90E-06&	0.0096&	1.95E-06&	0.0096&	 2.45E-06&	0.0096&	1.97E-06 \\ \hline
2100&	0.0097&	3.14E-06&	0.0097&	3.11E-06&	0.0097&	3.11E-06&	0.0098&	 3.39E-06&	0.0097&	3.04E-06 \\ \hline
2400&	0.0099&	1.32E-06&	0.0100&	1.34E-06&	0.0100&	1.35E-06&	0.0101&	 1.64E-06&	0.0100&	1.44E-06 \\ \hline
2700&	0.0100&	1.72E-06&	0.0100&	1.69E-06&	0.0100&	1.67E-06&	0.0101&	 2.11E-06&	0.0100&	1.90E-06 \\ \hline
3000&	0.0102&	2.96E-06&	0.0102&	2.93E-06&	0.0102&	2.91E-06&	0.0103&	 2.83E-06&	0.0102&	2.87E-06 \\ \hline
4800&	0.0103&	1.74E-06&	0.0104&	1.74E-06&	0.0104&	1.74E-06&	0.0104&	 2.06E-06&	0.0104&	2.01E-06 \\ \hline
9900&	0.0099&	8.18E-07&	0.0099&	8.30E-07&	0.0099&	8.36E-07&	0.0099&	 9.78E-07&	0.0099&	9.11E-07 \\ \hline
\end{tabular}
  \caption{Simulation Result of a 8-Descendant Tree, 6 of the 8 have Loss Rate=$1\%$ and the other 2 have Loss Rate=$5\%$}
  \label{Tab3}
\end{table*}

\begin{table*}
\centering
\scriptsize
\begin{tabular}{|l|r|r|r|r|r|r|r|r|r|r|}  \hline
 Estimators &\multicolumn{2}{|c|}{Full Likelihood} & \multicolumn{2}{|c|}{$A_k(2)$} & \multicolumn{2}{|c|}{$A_k(3)$} &\multicolumn{2}{|c|}{$Al_k(x), |x|=2$} & \multicolumn{2}{|c|}{$Al_k(x), |x|=3$}	\\ \hline
samples & Mean & Var &	Mean & Var &	Mean & Var	& Mean & Var &  Mean &	Var	\\ \hline
300&	0.0503&	2.15E-04&	0.0504&	2.15E-04&	0.0505&	2.14E-04&	0.0508&	2.18E-04&	0.0505&	2.16E-04\\ \hline
600&	0.0503&	8.23E-05&	0.0503&	8.21E-05&	0.0503&	8.19E-05&	0.0504&	8.24E-05&	0.0503&	8.27E-05\\ \hline
900&	0.0511&	5.85E-05&	0.0511&	5.81E-05&	0.0511&	5.79E-05&	0.0512&	5.79E-05&	0.0512&	5.88E-05\\ \hline
1200&	0.0506&	4.93E-05&	0.0506&	4.97E-05&	0.0507&	4.99E-05&	0.0507&	4.85E-05&	0.0507&	4.93E-05\\ \hline
1500&	0.0502&	2.24E-05&	0.0502&	2.24E-05&	0.0502&	2.23E-05&	0.0503&	2.33E-05&	0.0502&	2.32E-05\\ \hline
1800&	0.0500&	3.89E-05&	0.0500&	3.85E-05&	0.0500&	3.83E-05&	0.0501&	3.91E-05&	0.0500&	3.94E-05\\ \hline
2100&	0.0507&	1.16E-05&	0.0507&	1.19E-05&	0.0507&	1.20E-05&	0.0507&	1.09E-05&	0.0507&	1.13E-05\\ \hline
2400&	0.0510&	1.40E-05&	0.0510&	1.43E-05&	0.0510&	1.44E-05&	0.0510&	1.40E-05&	0.0510&	1.43E-05\\ \hline
2700&	0.0507&	1.31E-05&	0.0507&	1.34E-05&	0.0507&	1.35E-05&	0.0508&	1.35E-05&	0.0507&	1.34E-05\\ \hline
3000&	0.0508&	6.65E-06&	0.0508&	6.98E-06&	0.0508&	7.14E-06&	0.0508&	6.79E-06&	0.0508&	6.85E-06\\ \hline
4800&	0.0498&	1.09E-05&	0.0498&	1.10E-05&	0.0498&	1.10E-05&	0.0498&	1.11E-05&	0.0498&	1.11E-05\\ \hline
9900&	0.0496&	5.35E-06&	0.0496&	5.38E-06&	0.0497&	5.40E-06&	0.0496&	5.48E-06&	0.0496&	5.48E-06\\ \hline
\end{tabular}
  \caption{Simulation Result of a 8-Descendant Tree, the loss rate of the root link=$5\%$, 4 of the 8 have Loss Rate=$1\%$ and the other 4 have Loss Rate=$5\%$}
  \label{Tab4}
\end{table*}

\subsection{Simulation}

To compare the performance of the estimators between the original MLE, $A_k(i)$, and $Al_k(x)$, three rounds of simulations are conducted in various setting. Five estimators: the original MLE, $ A_k(2), A_k(3), Al_k(x), |x|=2$,  and $Al_k(x), |x|=3$, are compared against each other in the simulation and the results are presented in three tables, from Table \ref{Tab2} to Table \ref{Tab4}. The number of samples used in the simulations varies from 300 to 9900 in a step of 300. For each sample size, 20 experiments with different initial seeds are carried out and the means and variances of the estimates obtained by the five estimators are presented in the tables for comparison. Due to the space limitation, we only present a part of the results in the tables, where all of the means and variance for the samples varying from 300 to 3000 are included. For the samples from 3300 to 9900, only two of them, i.e. 4800 and 9900, are presented.

 Table \ref{Tab2} is the results obtained from a tree with 8 subtrees connected to node $k$, where the loss rate of the subtrees are set to 1\%.  The result shows when the sample is small, the estimates obtained by all estimators are drifted away from the true value that indicates the data obtained is not stable. Once the sample size reaches 2100, the estimates approach to the true value because the data is stabilised around the true value. All of the estimators achieve the same outcome with the increase of samples. Generally, with the increase of samples, the variance reduces slowly although there are a number of exceptions.  This indicates that there is no significant advantage of the original MLE and BWE over IBE if the subtrees connected to path of interest have the same loss rates. Therefore, by examining the pass rates of the paths connecting the source to the receivers of the subtrees, one is able to find the most suitable estimator.

To see the impact of different loss rates at the subtrees on estimates, another round simulation is carried out on the same network topology.
 The difference between this round and the previous one is the loss rates of the subtrees connected to the path of interest, where 6 of the 8 subtrees have their loss rates equal to $1\%$ and the other two have their loss rates equal to $5\%$. The two subtrees
 selected by the paired local estimator have their loss rates equal to $1\%$ and $5\%$, respectively. Two of the three subtrees used by $Al_k(x), |x|=3$ have their loss rates equal to $1\%$ and the other has its loss rate equal to $5\%$. The results are presented in Table \ref{Tab3}. Compared Table \ref{Tab3} with Table \ref{Tab2}, there is no change for the original MLE and there are slight changes for the estimators of the pairwise likelihood ($A_k(2)$) and the triplet-wise likelihood ($A_k(3)$). In contrast, the variances and the means of the other two  have noticeable differences from their counterparts, in particular  if the sample size is smaller than 1000.  In addition, the two have a  slightly higher variances than that obtained in the first round in general.
This indicates that the sensitivity of the estimators in IBE in terms of selecting observation or observers for estimation. If applying the result derived from (\ref{Al variance1}), we should  examine  the pass rates of the paths connecting the source to the subtrees first. Then,   the subtrees that have loss rates equating to $1\%$ would be selected for $Al_k(x), |x| = 2$ or $3$. If so, the same result as the most right four columns of table \ref{Tab2} will be obtained that is certainly better than that in table \ref{Tab3}.

To further investigate the impact of loss rates on estimation, we  conduct the third round simulation, where the loss rate of the path of interest is increased from $1\%$ to $5\%$, and the loss rates of four subtrees are set to $5\%$ and the other four to $1\%$.  The two estimators from IBE, i.e. $Al_k(x), |x|=2$ and $3$, consider the observations obtained from the subtrees that have $1\%$ loss rate. The result is presented in Table \ref{Tab4}, it differs from the previous two tables in the estimated variances that are a magnitude higher than that of the previous two regardless of the estimators. This is actually an expected result of (\ref{Al variance1}),  i.e., a smaller $\delta_k(x)$ results in a bigger variance.

\section{Conclusion}\label{section7}

This paper starts from finding inspirations that can lead to efficient explicit estimators for loss tomography and ends with a large number of unbiased and consistent explicit estimators, plus a number of theorems and corollaries to assure the statistical properties of the estimators. One of the most important findings is of the formulae to compute the variances of $A_k$ estimated by the estimators in RSE, IBE and the original MLE. Apart from clearly expressing the connection between the path to be estimated and the subtrees connecting the path to the observers of interest, the formulae potentially have many applications in network tomography, some have been identified in this paper.  For instance, using the formulae, we have ranked the MLEs proposed so far, including those proposed  in this paper. In addition, the formulae make model selection possible in loss tomography and then the multicast used in end-to-end measurement is no longer only for creating various correlations but also for identifying the subtrees that can be used in estimation. The effectiveness of the strategy has been verified in a simulation study. Despite this, the potentials of the formulae have not reached that require further exploration.


\bibliography{congestion}

\begin{thebibliography}{10}

\bibitem{DHPT06}
N.~Duffield, J.~Horowitz, F.~L. Presti, and D.~Towsley, ``Explicit loss
  inference in multicast tomography,'' {\em IEEE trans. on Information Theory},
  vol.~52, no.~8, 2006.

\bibitem{CDHT99}
R.~C\'{a}ceres, N.~Duffield, J.~Horowitz, and D.~Towsley, ``Multicast-based
  inference of network-internal loss characteristics,'' {\em IEEE Trans. on
  Information Theory}, vol.~45, 1999.

\bibitem{CDMT99}
R.~C\'{a}ceres, N.~Duffield, S.~Moon, and D.~Towsley, ``{Inference of Internal
  Loss Rates in the MBone },'' in {\em IEEE/ISOC Global Internet'99}, 1999.

\bibitem{CDMT99a}
R.~C\'{a}ceres, N.~Duffield, S.~Moon, and D.~Towsley, ``Inferring link-level
  performance from end-to-end multicast measurements,'' tech. rep., University
  of Massachusetts, 1999.

\bibitem{CN00}
M.~Coates and R.~Nowak, ``Unicast network tomography using {EM} algorthms,''
  Tech. Rep. TR-0004, Rice University, September 2000.

\bibitem{XGN06}
B.~Xi, G.~Nichailidis, and V.~Nair, ``Estimating netwrok loss rates using
  active tomography,'' {\em JASA}, 2006.

\bibitem{BDPT02}
T.~Bu, N.~Duffield, F.~Presti, and D.~Towsley, ``Network tomography on {General
  Topologies},'' in {\em SIGCOMM 2002}, 2002.

\bibitem{ADV07}
V.~Arya, N.~Duffield, and D.~Veitch, ``Multicast inference of temporal loss
  charateristics,'' {\em Performance Evaluation}, vol.~9-12, 2007.

\bibitem{ZG05}
W.~Zhu and Z.~Geng, ``Bottom up inference of loss rate,'' {\em Journal of
  Computer Communications}, vol.~28, no.~4, 2005.

\bibitem{GW03}
D.~Guo and X.~Wang, ``Bayesian inference of network loss and delay
  characteristics with applications to tcp performance predication,'' {\em IEEE
  trans. on Signal Processing}, vol.~51, no.~8, 2003.

\bibitem{Zhu11a}
W.~Zhu, ``An efficient loss rate estimator in multicast tomography and its
  validity,'' in {\em IEEE International Conference on Communicaation and
  Software}, 2011.

\bibitem{Lindsay88}
B.~C. Lindsay, ``Composite likelihood method,'' {\em Contemporary Mathematics},
  vol.~80, 1988.

\bibitem{YV96}
Y.~Vardi, ``Network tomography: Estimating source-destination traffic
  intensities from link data,'' {\em Journal of Amer. Stat. Association},
  vol.~91, no.~433, 1996.

\bibitem{HBB00}
K.~Harfoush, A.~Bestavros, and J.~Byers, ``Robust identification of shared
  losses using end-to-end unicast probes,'' in {\em Technical Report
  BUCS-2000-013}, Boston University, 2000.

\bibitem{RM96}
R.~Mittelhammer, {\em Mathematical Statistics for Economics and Business},
  vol.~78.
\newblock Springer, 1996.

\bibitem{Besay74}
J.~Besag, ``Spatial interaction and the statistical analysis of lattice system
  (with discussion),'' {\em Journal of Royal Statistical Society}, vol.~36,
  1974.

\bibitem{Joshi76}
V.~M. Joshi, ``On the attainment of the cramer-rao lower bound,'' {\em Ann.
  Statist.}, vol.~4, no.~5, pp.~998--1002, 1976.

\end{thebibliography}


\end{document}